# The Primordial Perturbation Spectrum and Large Scale Structure

STEFAN GOTTLÖBER

*Astrophysikalisches Institut Potsdam*
*An der Sternwarte 16*
*D-14482 Potsdam*
*sgottloeber@aip.de*

## INTRODUCTION

The standard model of structure formation is based on two assumptions, namely, that the primordial perturbation spectrum generated during inflation is of Harrison-Zeldovich type and, that the dark matter, which gives the main contribution to the density of the Einstein-deSitter universe, is cold. Recent observations have shown that this model cannot explain all observational facts together[1-4]. In order to change the theoretical predictions of the standard model, one can modify either of these assumptions. Widely discussed candidates for closing the universe are mixed dark matter[5] or a cosmological term[6]. Due to the additional matter component the evolution of the perturbations is changed so that the resulting spectrum differs from the standard one. Improved inflationary scenarios lead to other than Harrison-Zeldovich spectra[7,8] or even to an open cosmological model[9]. Changing the primordial perturbation spectrum one can maintain both the successful description of structure formation over a wide range of scales by the CDM model and the Einstein-deSitter model predicted by almost all inflationary scenarios. Our underlying inflationary model has two consecutive stages of exponential expansion with a short intermediate stage of power law expansion[7]. In consequence of the intermediate stage the scale invariance of the resulting perturbation spectra is broken, i.e. they are of Harrison-Zeldovich type only in the limit of very small and very large scales. In the intermediate range the power is scale dependent. Besides the normalization the BSI (Broken Scale Invariance) spectra are characterized by the ratio $\Delta$ of the power on large scales to that on small scales, and a scale $k_{br}^{-1}$ denoting the onset of the break in the perturbation spectrum at small scales.

## OBSERVATIONAL CONSTRAINTS

At large scales we have to normalize the resulting power spectra by means of the COBE data. There are different approaches to this procedure. We are using the normalization proposed by Górski et al.[10] who calculated the multipole $a_9$ of the CMB temperature fluctuations $\Delta T/T$ from the COBE data. In this case the normalization of the spectra is independent of the spectral index which is for our spectra slightly different from 1 on COBE scales (cp. Fig. 1 and 2). We have calculated the multipole moment of the CMB background fluctuations for different primordial perturbation spectra[11]. Our approximation agrees within a few percent with the approximation proposed by Naoshi Sugiyama (see this proceedings), in particular it gives exactly the same result for low multipoles which are relevant for the normalization. Normalizing the perturbation spectra with the COBE data we find the biasing parameter $b = \sigma_8^{-1} \approx 1.7$. In Fig. 1 the BSI spectrum is shown in comparison with a Harrison-Zeldovich spectrum (both for $\Omega = 1, H = 50$ km/s/Mpc) and a spectrum for a model with cosmological term ($\Omega + \lambda = 1, \lambda = 0.8$).

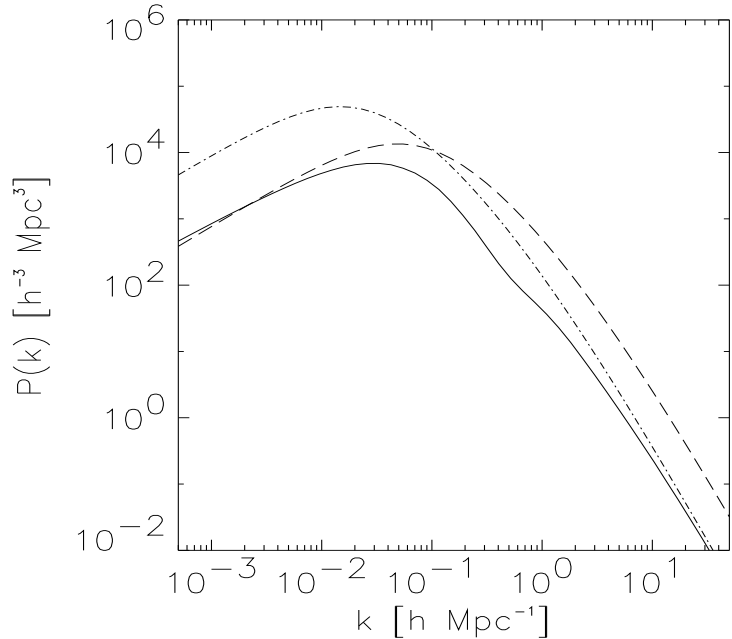

**Figure 1:** The linear perturbation spectra of the BSI model (solid line), the standard CDM model (dashed line), and a model with a cosmological term (dash-dotted line).

We have compared the theoretical predictions of different BSI spectra with such observational data as the variances calculated from the counts in cells, the angular correlation function, the Mach numbers, or the rms mass fluctuations necessary for quasar and galaxy formation[12]. These data appear to favour a spectrum with $\Delta = 3$ and an onset of extra power at about 2 $h^{-1}$Mpc. In that case the Hubble parameter is assumed to be 50 km/s/Mpc. On the other hand, recent observations with the HST suggest a big Hubble parameter of about 75 km/s/Mpc or even higher (see the contributions of W. Freedman and T. Shanks in these proceedings). Such a high Hubble parameter requires a nonzero cosmological term. This changes the predicted multipoles of the CMB fluctuations. In Fig. 2 the multipoles for the Harrison-Zeldovich and BSI spectra in a CDM model with $H = 50$ km/s/Mpc are shown in comparison with a model with $\lambda = 0.8$, $\Omega = 0.2$, $H = 75$ km/s/Mpc.

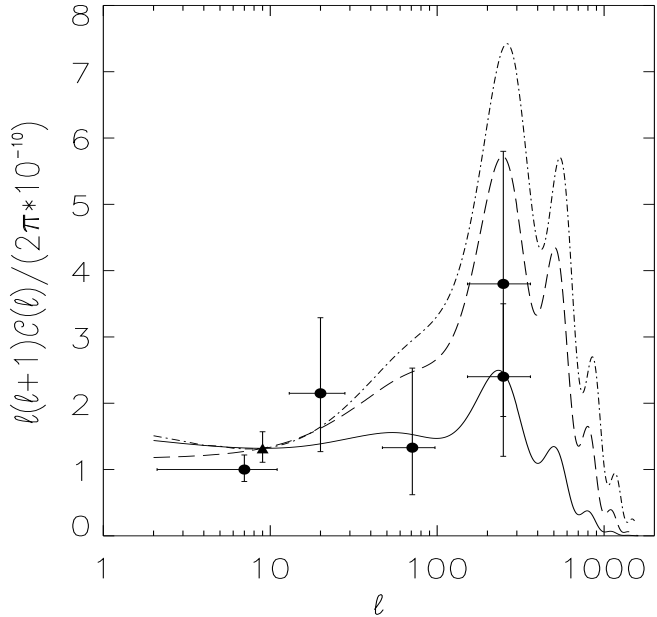

**Figure 2:** The multipole moments of the fluctuations of the CMB radiation for the BSI model (solid line), the standard CDM model (dashed line), and a model with a cosmological term (dash-dotted line).

The triangle in Fig. 2 denotes the Górski normalization. The left experiment is the slightly lower $\sigma_{10}$ COBE measurement. It follow from left to right the Teneriffe, SK93 and MSAM3 (full and source free) data points (courtesy B. Ratra, from Bond[13]). The data points and error bars are an illustration. One has to calculate the predicted $\Delta T/T$ of each model by convolving the multipole moments with the filter function of the experiment in order to compare models correctly. As one can see easily from Fig. 2 models with a cosmological term predict a much higher first Doppler peak than the standard model. On the other side BSI models predict a lower Doppler peak. Thus better measurements of the multipoles around 100 will become a crucial test to discriminate between these models.

## N-body simulations

Using the BSI spectrum shown in Fig. 1 and a Harrison-Zeldovich spectrum we have performed N-body simulations and calculated from these simulations galaxy correlation functions and higher moments, the cluster abundances, reconstructed power spectra for the density fields, the angular correlation function and further statistics[14,15]. The simulations show that the model with broken scale invariance fits the observational data quite well also in the nonlinear regime. In particular, the BSI model yields the right slope of the APM angular correlation function over all simulated angular scales, whereas the standard CDM model shows a too steep slope[14]. The observed cluster mass function is fitted very well by the BSI models.

The slope of the mass function in the standard CDM model is too steep on cluster scales[15]. In the BSI model the structure formation occurs later than in the CDM model with a Harrison-Zeldovich spectrum, i.e. the standard model shows to much structures at $z = 0$.

Obviously, one uses only a small part of the spectrum in the simulations. From Fig. 1 it is clear that for small boxes ($l_{Box} < 25h^{-1}$ Mpc) with decreasing box sizes the simulations become more and more equivalent to a CDM model with Harrison-Zeldovich spectrum and a normalization of $\sigma_8 \approx b_{lin}^{-1} = 0.6$. The simple reason is that simulations in these boxes feel only the reduced power produced during the second inflationary stage (the slope of the BSI spectrum approximately coincides with the Harrison-Zeldovich spectrum). Thus, the promising results obtained in simulations at low normalization ($\sigma_8$ between 0.6 and 0.8; note however, that the COBE normalized Harrison-Zeldovich spectrum with CDM yields $\sigma_8 > 1$.) of small scale structure within the CDM model are preserved in the BSI model.

## CONCLUSION

Based on confrontation with observations and comparison with the COBE-normalized CDM model, we conclude that the perturbation spectra with broken scale invariance (BSI models) represent a promising modification of the CDM picture. Under the standard assumption of an Einstein-deSitter model ($\Omega = 1$) with $\Omega_{bar} = 0.06$ the BSI models allow to describe many details of large scale structure formation.